# Resonant Rayleigh scattering from quantum phases of cold electrons in semiconductor heterostructures

S. Luin[1,3], V. Pellegrini[1], A. Pinczuk[2,3], B.S. Dennis[3], L.N. Pfeiffer[3], K.W. West[3]

1. NEST CNR-INFM and Scuola Normale Superiore, Pisa (Italy)
2. Dept. of Appl. Phys. & Appl. Math., and Dept. of Physics, Columbia University (NY)
3. Bell Labs, Lucent Technologies Murray Hill (NJ)

Resonant Rayleigh scattering of light from electrons confined in gallium arsenide double quantum wells displays significant changes at temperatures that are below one degree Kelvin. The Rayleigh resonance occurs for photon energies that overlap a quantum well exciton and when electron bilayers condense into a quantum-Hall state. Marked changes in Rayleigh scattering intensities that occur in response to application of an in-plane magnetic field indicate that the unexpected temperature dependence is linked to formation of non-uniform electron fluids in a disordered quantum-Hall phase. These results demonstrate a new realm of study in which resonant Rayleigh scattering methods probe quantum phases of cold electrons in semiconductor heterostructures.

The elastic scattering of radiation (photons, neutrons, etc.) is an elementary process that was described by Lord Rayleigh for sunlight radiation in the presence of fluctuations of the atmosphere's index of refraction [1]. Elastic (Rayleigh) scattering occurs in inhomogeneous condensed matter systems and in crystals with defects. Rayleigh scattering of light has found remarkable application in the study of phase transitions in gases and liquids [2]. Scaling phenomena eventually lead to the critical opalescence that was first documented at the critical point of carbon dioxide [3].

More than twenty years ago Hegarty et al. reported Rayleigh light scattering in *GaAs/AlGaAs* quantum wells at temperatures of the order of several degrees Kelvin [4]. The elastic scattering intensities from quantum wells display strong enhancements when photon energies overlap those of interband excitonic transitions. The resonant Rayleigh scattering from the semiconductor quantum wells has been explained by inhomogeneous fluctuations in quantum well width that cause the localization of optical excitons [4,5,6].

Here we report that resonant Rayleigh scattering intensities from *GaAs/AlGaAs* quantum structures display unexpected temperature dependence for changes that occur in the milli-Kelvin range. The effect is observed in double quantum wells that support two coupled two-dimensional (2D) electron gases in the quantum Hall (QH) regime [7]. Examples of the novel behavior are seen in Figs. 2a and 3a, for bilayers in the QH state with Landau level filling factor $\nu=1$, where $\nu=2\pi n l_B^2$, $l_B^2 = \hbar c/eB$, $n$ is the surface electron density and $B$ the perpendicular magnetic field. The photon energy of the maximum enhancement of resonant Rayleigh scattering overlaps that of optical transitions in the quantum well structures. The resonance is extremely narrow. The width of only about 0.2 meV (~1Å) reveals the exciton character of the resonance. These measurements reveal that the large, resonantly enhanced, Rayleigh scattering intensities seen at 60 milli-Kelvin degrees disappear as the temperature approaches one degree Kelvin.

The resonant Rayleigh scattering (RRS) in electron bilayers at $\nu=1$ occurs when the cold liquid is in QH phases with strong inter-layer electron correlation that were reported in Ref. [8]. These quantum phase transformations were discovered in the anomalous behavior of low-lying spin excitations (seeFig.1a). These are quantum states that display electron occupation of Landau levels that depart markedly from the conventional filling in which only the lowest Landau level is populated. The characteristic signature of these quantum phases of the bilayers is found here in the marked dependence of RRS on the strengths of in-plane components of magnetic field at low temperatures [8].

The link between RRS and the quantum phase transition reported in Ref. [8] is exemplified by the results in Fig. 3c that reveal that the temperature dependence of RRS below 1K is tuned by application of an in-plane component of magnetic field. The observation of strong RRS reported here offers evidence that these QH fluids of electron bilayers at $\nu=1$ are highly non-uniform showing domains with characteristic lengths that are comparable to the wavelength of light in the medium (~2000Å). The physics of the correlated QH phases that emerge at $\nu=1$ manifests in the specific temperature and magnetic field dependences of RRS. We recall that non-uniform QH states are known to occur in single electron layers near QH states in which weak residual disorder leads to sub-micron sized regions with values of $\nu$ that deviate from the exactly quantized value [9,10]. Separation into non-uniform phases was recently predicted to occur in electrons bilayers at $\nu=1$ [11-14] and also in low-density 2D electronic system close to the transition to the Wigner crystal [15]. The resonant Rayleigh scattering results uncover a new domain of applications in which RRS effects probe quantum phases of electrons that emerge at low temperatures in low-dimensional semiconductor structures.

Figure 1 describes the experimental set-up. Samples are kept in a dilution refrigerator with windows for optical access. Its base temperature is below 50 mK. The perpendicular and in-plane components of magnetic field are $B=B_T\cos\theta$ and $B_{//}=B_T\sin\theta$, where $B_T$ is the total field. B was always chosen to correspond to $\nu=1$ for all values of tilt angle $\theta$. The photon energies of the emission of dye, diode or titanium-sapphire lasers were tuned to be close to the fundamental interband optical gap of the double quantum well ($\lambda \approx 810$ nm). The in-plane wave-vector transfer in the light scattering in the back-scattering geometry employed here is given by $q=|\mathbf{k}^{//}_S-\mathbf{k}^{//}_L|= 4\pi\sin(\theta)/\lambda$, where $\mathbf{k}^{//}_{S(L)}$ is the in-plane projection of the wavevector of the scattered (laser) light. The non-specular scattered light was collected into a double or triple spectrometer equipped with a CCD detector with spectral resolution of 15 μeV. A crossed polarization scheme with incident laser polarization perpendicular to the scattered light polarization was adopted in order to reduce the impact of the stray laser light collected by the spectrometer. Optimization of the optical set-up was crucial for the observation of the RRS signal.

Two modulation-doped $Al_{0.1}Ga_{0.9}As/GaAs$ double quantum wells were grown by molecular beam epitaxy. They have total electron densities of $n = 1.2\times10^{11}$ cm$^{-2}$ (sample 1) and $n = 1.1\times10^{11}$ cm$^{-2}$ (sample 2). The spacing $d$ between quantum wells has $d/l_B$ values of 2 and 2.2 respectively. The values of the tunneling gaps $\Delta_{SAS}$ are *0.58* meV and *0.36* meV [16]. A schematic representation of the conduction-band profile and lowest electron levels at $v=1$ is shown in Fig.1(a), where S and A label the symmetric and anti-symmetric linear combinations of spin-split Landau levels of left and right quantum well. The lowest-lying spin excitations at $v=1$, that are detected together with RRS, are the spin-flip (SF) mode across $\Delta_{SAS}$ with simultaneous change of spin and, at lower energy, the spin-wave (SW) mode across the Zeeman gap [8]. The transitions that build SW and SF modes at $v=1$ are also shown in Fig. 1a.

Figure 2 displays a resonant enhancement profile of Rayleigh scattering intensity recorded in sample 1. Measurements are performed at an incident laser intensity of $\approx 10^{-4}$ W/cm$^2$ and $\theta = 30°$. The Rayleigh scattering peaks have line-width limited by the resolution of the spectrometer. In Fig.2a the Rayleigh scattering peaks are plotted for different incident laser wavelengths and compared to the broad luminescence profile of the $v=1$ QH fluid. The sharp resonant profile of the Rayleigh intensity (FWHM of approximately 1-2 Å ≈ 0.2-0.4 meV) and the photon energy of maximum enhancement located on the high-energy side of the luminescence indicate that the Rayleigh scattering relies on an inter-band excitonic resonance of the semiconductor heterostructures. In this exciton the electron states are in the first excited (spin-down) symmetric Landau level shown in Fig.1b.

The color plot in Fig.2b reports the intensity of light scattering as a function of the incident laser wavelength (horizontal axis) and of the energy shift of the scattered light from the photon energy of the incoming laser light (vertical axis). In this plot, the elastic resonant Rayleigh component of the scattered radiation occurs at zero energy shifts. At finite positive energy shifts we also observe an optical recombination emission due to magneto-luminescence and the inelastic light scattering signals due to SF and SW excitations. The energy splitting between these two excitations is much smaller than the value of $\Delta_{SAS}$ = 0.58 meV that could be expected from the picture shown in Fig.1a and from mean-field theories [17]. In recent works it was shown that the reduction of the energy splitting implies that the QH ground state is a highly-correlated phase that departs from conventional QH (it incorporates excitonic electron-hole pairs across the tunneling gap) [8,18].

Figure 3 shows the marked temperature dependence of the RRS intensity measured in sample 2. At the lowest tilt angle of $\theta=5^o$ the RRS disappears as the temperature is raised above $T$=0.8 K (Fig.3a and top curve in Fig.3c). The weak Rayleigh scattering that remains at high temperature (above 1.3K) can be linked to the impact of residual disorder mechanisms also at work in undoped quantum wells [4-6] that however in our high-mobility modulation-doped samples are suppressed significantly. When $\Delta_{SAS}$ is reduced in sample 2 by increasing the in-plane component of the magnetic field (keeping the perpendicular field to the $\nu=1$ value) the temperature evolution becomes more pronounced with the RRS disappearing just above 0.4 K at $\theta$ = 30°. The suppression of $\Delta_{SAS}$ by changing the tilt angle is given by $\Delta_{SAS}(\theta)$ = $\Delta_{SAS}(\theta=0)exp(-d/l_B tg^2\theta)$ [19]. As θ becomes equal or larger than the critical value of $\theta_c \approx 35^o$

($\Delta_{SAS}/E_C$ = 0.018) both the RRS intensity and its resonant profile and temperature behavior change abruptly signaling the occurrence of a phase instability.

The dependence of the RRS intensity on temperature and $B_{//}$ shown in Fig. 3 are nearly identical to those of spin excitations that are characteristic signatures of highly correlated QH phases of the electron bilayers [8]. The results in Fig. 3 thus establish an unambiguous link between RRS and the phase transformations of the correlated QH fluids in the electron bilayers. The results shown in Figs. 4a and 4b, that compare the RRS at representative angles below and above $\theta_c$, provides key evidence on the nature of the quantum phase that emerges above $\theta_c$: the vanishing of the resonant component of the Rayleigh scattering intensity for angles $\theta \geq \theta_c$ indicates that the emergent phase is significantly more homogeneous than the bilayer QH phase that occurs for $\theta < \theta_c$.

To construct a quantitative description of the RRS intensity we assume that the non-uniform phase consists of two types of domains that have different frequency-dependent dielectric constants associated with two characteristic lengths $\xi_1$ and $\xi_2$, and average distance $\xi_1 + \xi_2$ [5,20]. The $\theta$-dependence of the RRS intensity requires changes in $\xi_1$ and $\xi_2$ that occur when $B_{//}$ is changed. The maximum RRS intensity in the far field is obtained when $\xi_1$ and $\xi_2$ are comparable and close to $\lambda/2\pi sin\theta$ [20]. To identify the nature of the domains and interpret the temperature dependence we recall that previous experiments [8,16] have shown that in the range of the $\Delta_{SAS}$ and $d/l_B$ values of our samples and below $\theta_c$ the bilayer electron fluid at $\nu=1$ is in the QH phase. This indicates that the dominant domains stable at low-temperature are in the correlated QH state described in Ref. [8] and create a percolation path throughout the sample allowing charge transport consistent with the QH effect. In Ref. [8] we showed that a fraction of electrons occupies the higher-energy spin-up anti-symmetric Landau level in this correlated phase. The occurrence of the QH effect at $\nu=1$ thus implies those electrons to be bound to the holes left behind in the lowest-energy symmetric Landau level to create neutral excitonic-like particles not contributing to charge transport.

The temperature dependence of the RRS intensity is linked to changes in the two types of domains. We can argue that as temperature increases the QH domains composed by the correlated excitonic phase dissolve into the other phase reducing the spatial variation of the dielectric functions. Such temperature related changes will tend to suppress the RRS intensity. We show below that the thermodynamic equilibrium for the two phases and thus the RRS intensity is well described by the 2D version of the Langmuir adsorption isotherm. This isotherm, originally introduced to account for the equilibrium established between molecules in the gas phase and the corresponding adsorbed species bound to binding sites on the surface of a solid, was applied to describe thermally induced exciton de-localization in quantum wells [21].

In this framework the *T*-dependence of the resonant Rayleigh intensity $I_{RRS}$ is given by:

$$I_{RRS}(T) = \frac{I^o_{RRS}}{1 + C \cdot T \exp(-E_b/kT)} \qquad (1)$$

In Eq. (1) $E_b$ is a binding energy required to destroy the excitonic correlation in the individual domains. One possible mechanism suggested by the Langmuir behavior could involve unbinding of mobile electrons into the correlated phase. These processes reduce the mismatch of the dielectric constants among the two domains suppressing the RRS. Within this interpretation $C=2\pi m_e k/N_p h^2$ and $N_p$ can be viewed as the density of binding sites. The solid black line on the top of Fig.3c represents the best-fit of the experimental RRS intensities to Eq.1. at the tilt angle of 5º. The best fit analysis yields $E_b$=0.27 meV (3.3K) and $C$= 0.2 K$^{-1}$ ($N_p$ = 0.6x10$^{10}$ cm$^{-2}$ assuming for simplicity $m_e$=0.067$m_o$ where $m_o$ is the free electron mass). The adjusted value of $E_b$ is similar to that found in thermo-activated transport experiments in QH bilayers at *v=1* [22] and interpreted in terms of a finite-temperature phase transition twowards a non-correlated state. It is intriguing to link this transition to a cooperative phenomenon of ionization of the excitons reported in Ref [8]. To this end we note that typical excitonic binding energies calculated for our sample parameters within a time-dependent Hartree-Fock approach are consistent with the value of $E_b$=0.27 meV [17,23].

The temperature dependences can be fitted by Eq.2 at all the investigated tilt angles (below and above 35º). However while below $\theta_c$ the values of C range between C=0.2K$^{-1}$ (5º) and C=3 K$^{-1}$ (30º), at angles of $\theta_c$=35º or above the fit yields C=0.006 K$^{-1}$ or less signaling the emergence of a different (more homogeneous) physical configuration. Solid lines in Fig.3c represent the best-fit to Eq.2 at different angles. Figures 4c,d,e report the values of the binding energy $E_b$, $I^o{}_{RRS}$ and RRS linewidth versus angle or versus $\Delta_{SAS}(\theta)/E_c$. The phase transition at $\theta_c$ is signaled by a smooth decrease of the binding energy and abrupt changes of $I^o{}_{RRS}$ and RRS linewidth. Figure 4d also reports the RRS intensity values at 1.3K. At this temperature the effect associated to the structure of the electron liquid disappears and the RRS displays a marginal resonant effect and angle dependence. Figure 4d highlights the absence of a significant temperature-dependence of RRS in the phase above $\theta_c$.

A candidate state for the competing minority domains at low-temperature for $\theta<\theta_c$ is a phase with no intersubband excitons. This state at ν=1 is compressible and thus does not support the QH effect. In recent theoretical works a possible compressible phase at ν=1 was identified with a state composed by two weakly-interacting composite-fermion systems (each of them at ν=½) that is stabilized by intra-layer correlations [13]. It is possible that spontaneous phase separation occurs in the vicinity of the phase instability and that at $\theta>\theta_c$ the critical proliferation of such domains leads to a more homogeneous compressible fluid with marginal RRS. The transition to a compressible phase above $\theta_c$ is also consistent with the results of Ref. [8] and with the evolution of the position of our samples in the bilayer phase diagram [16] as the tunneling gap is suppressed. It is also worth noting that a non-uniform QH fluid is created by the impact of weak residual disorder [9,10] leading to coexistence of QH domains at $\nu=1$ with non-QH domains at $\nu \neq 1$. In the case of $\nu=1$ bilayers and above $\theta>\theta_c$ however, even the $\nu=1$ domains should not be in the QH phase and this would create a more homogeneous fluid suppressing the RRS.

Non-uniform electron liquids induced by weak residual disorder or by spontaneous phase separation are pervasive in the physics of correlated materials and low-dimensional structures. Stripe phases [24], ferromagnetic domain structures [25] and disorder-induced localized states [9] are prominent examples in the QH regime. The resonant Rayleigh scattering effect reported here opens new venues for the study of electron liquid phases that emerge at low temperatures in systems of reduced dimensionality.

We are grateful to Steven H. Simon for critical reading of the manuscript. V.P. acknowledges the support of the Italian Ministry of Foreign Affairs, the Italian Ministry of Research and by the European Community Human Potential Programme (Project HPRN-CT-2002-00291). A.P. acknowledges the National Science Foundation under Award Number DMR-03-52738, the Department of Energy under award DE-AIO2-04ER46133, the Nanoscale Science and Engineering Initiative of the National Science Foundation under Award Number CHE-0117752, and a research grant of the W.M. Keck Foundation.

Gaussian distribution for the dimensions of the two domains, is given by: $I_{RRS}^{o} \propto \Delta\varepsilon^2 \frac{\xi_1^2 \xi_2^2}{\left(\xi_1^2 + \xi_2^2\right)^3} \left( \xi_1^4 e^{-\frac{q^2 \xi_1^2}{4}} + \xi_2^4 e^{-\frac{q^2 \xi_2^2}{4}} \right)$ where $q$ is the in-plane wave-vector transfer in the light scattering process in the backscattering configuration and $\Delta\varepsilon$ is difference of the frequency-dependent dielectric constants in the two domains.

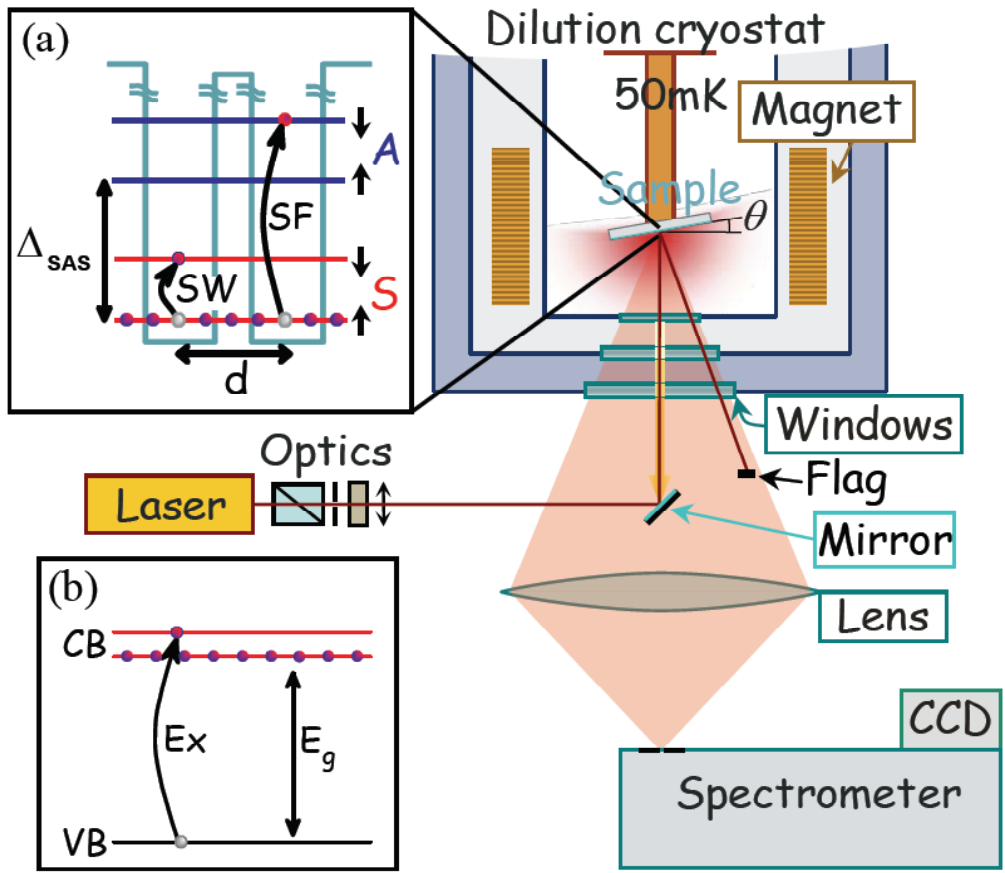

Figure 1. Schematic rendition of the experimental set-up used for the Resonant Rayleigh scattering measurements. A polarization rotator and two focusing lenses are the optical components after the laser. The lenses are used to create an elongated focus on the sample to match the rectangular slit aperture at the entrance of the spectrometer. A black flag is used to block the reflected laser light at low tilt angles θ. At larger tilt angles (θ≥10°), the reflected laser light is out of the collection cone (in red) and it is blocked by the internal wall of the dilution fridge. (a) Schematic view of the double quantum well in the single-particle-like conventional configuration at filling factor ν=1. Transitions in the spin wave (SW) and spin-flip (SF) modes are indicated by curved vertical arrows. Short vertical arrows indicate the orientation of spins, S and A label the symmetric and antisymmetric states separated by the tunneling gap $\Delta_{SAS}$. (b) Transitions that contribute to the interband magneto-exciton (Ex) responsible for the resonance in the Rayleigh scattering intensity. The red lines represent the spin up/down symmetric Landau levels. VB and CB indicate valence and conduction bands separated by the interband gap $E_g$.

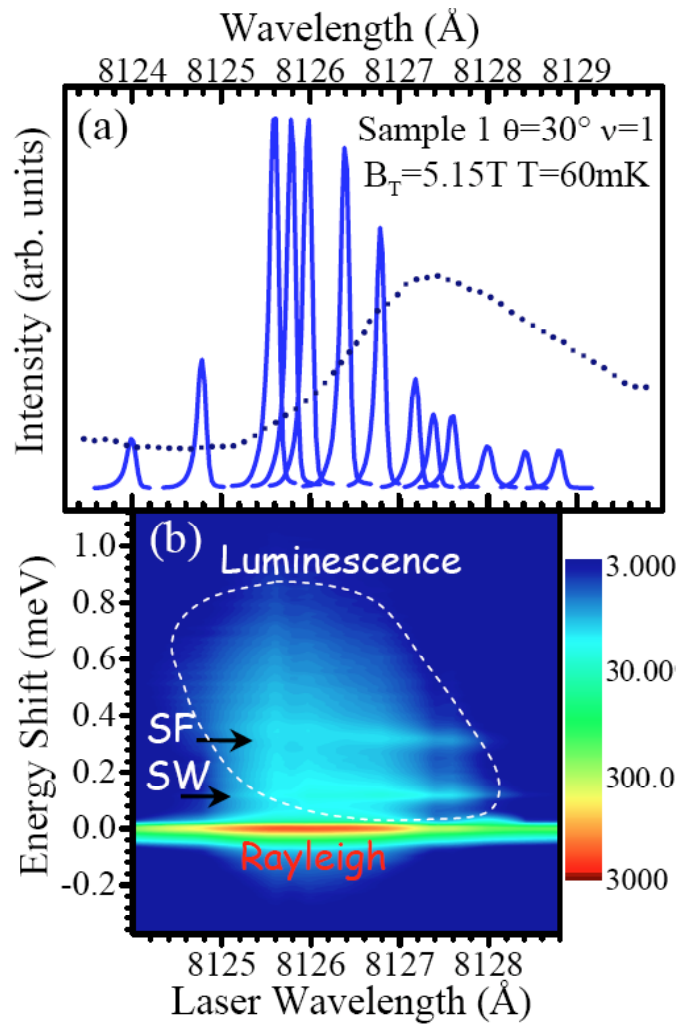

Figure 2. (a) Representative example of a resonant profile of the Rayleigh scattering intensity at $\nu=1$ and T=60 mK: each sharp peak (solid line) is the spectrum of elastically scattered light, with wavelengths and lineshapes that match exactly the ones of the incoming laser light. The dotted line is the magneto-luminescence with off-resonant excitation. (b) Logarithmic scale color plot of the light scattering intensities for the case reported in panel (a). The horizontal scale is the incident laser wavelength and the vertical scale is the energy shift between the scattered and incident light. The resonant Rayleigh scattering occurs at zero energy shift. Also indicated are the SW and SF inelastic light scattering peaks and the incoherent emission due to luminescence.

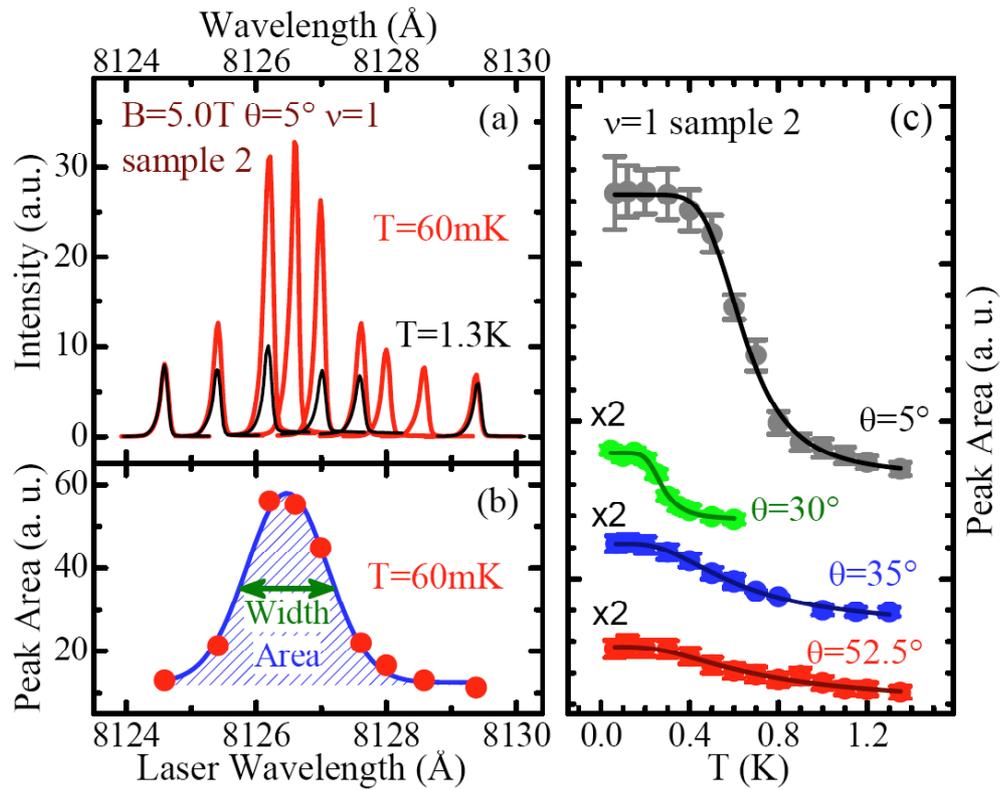

Figure 3. Resonant profile and temperature dependence of Rayleigh scattering for sample 2. (a) Evolution of the Rayleigh scattering peaks as a function of incident laser wavelength at two different temperatures. (b) Resonant profile for the Rayleigh scattering at T=60mK. (c) Temperature behavior of the resonant Rayleigh scattering at different angles. The best fits to Eq.(1) reported in the text are shown as solid lines. Data at each angle are shifted vertically for clarity.

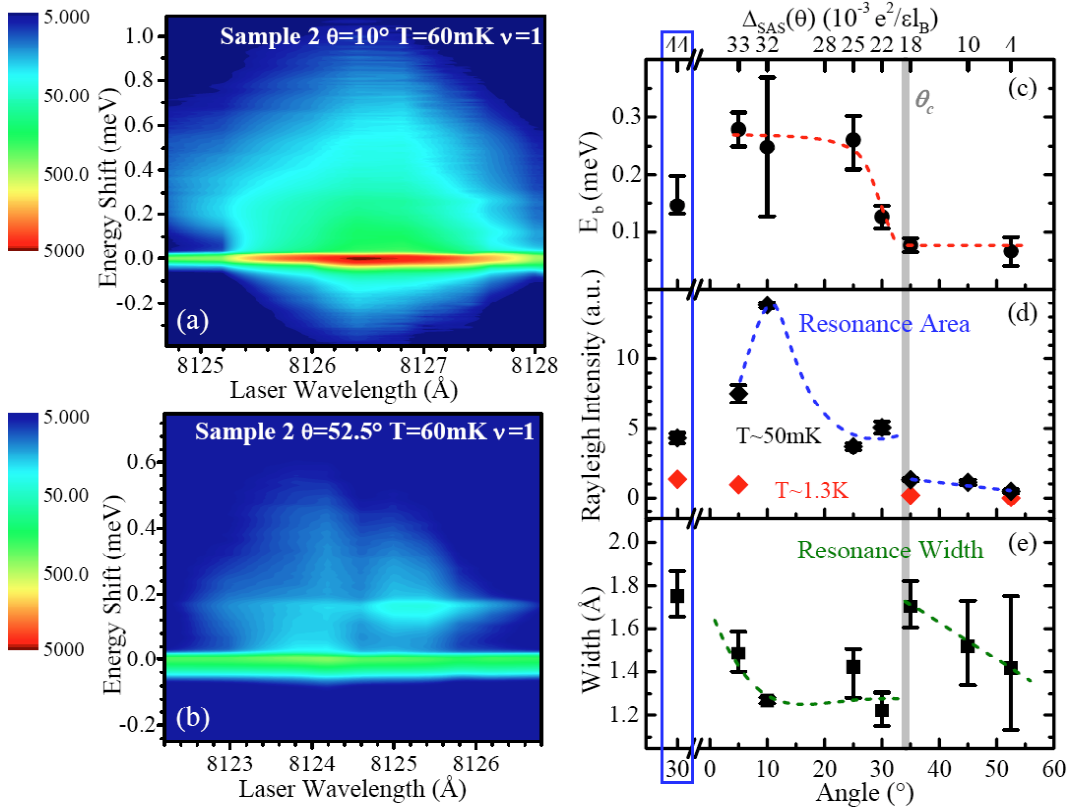

Figure 4. Evolution of the resonant Rayleigh scattering at the phase transition. (a) and (b) show color plots of the Rayleigh scattering at T=60mK. In (a) θ>θ$_c$. In (b) θ<θ$_c$, and only the spin wave inelastic peak can be seen at 0.18 meV. The sharp resonant enhancement of RRS observed in (a) and disappears above the critical angle of θ$_c$=35°. (c) Evolution of the binding energy E$_b$ (defined in Eq.2) as a function of angle (bottom horizontal axis) and $\Delta_{SAS}/(e^2/\varepsilon l_b)$ (top horizontal axis). Data are in sample 2 except for the points in the blue rectangle that refer to sample 1 at a tilt angle of 30°. (d) (e) same as in (c) but for the Rayleigh intensity (d) and resonance width (e) as obtained as in Fig.2b.